\documentstyle{article}

\begin{document}
\def \ts {\textstyle}
\def \rd {\displaystyle{\cdot}}
\def \ep {\varepsilon}
\def \p {\partial}
\def \cs {c_{\rm s}^2}
\def \b {\begin{equation}}
\def \e {\end{equation}}
\def \d {\mbox{d}}
\def \n {\noindent}
\def \l {\label}

\title{Exact solutions of Einstein's equations with ideal gas sources}  
\author{Roberto A.
Sussman \thanks{ email: sussman@nuclecu.unam.mx} \\
Instituto de Ciencias Nucleares. \\
Universidad Nacional Aut\'onoma de M\'exico.\\ A.P. 70-543, M\'exico D.F. 04510, 
M\'exico\\ \\ Josep Triginer\thanks{
email: pep@ulises.uab.es}\\
Departament de F\'{\i}sica. Facultat de Ci\`encies.\\ Universitat Aut\`onoma  de
Barcelona.\\ 08193 Bellaterra (Barcelona) Spain}
\date{\today}
\maketitle 

\abstract
{We derive a new class of exact solutions characterized by the Szekeres-Szafron metrics 
(of class I), admitting in general no isometries. The source is a fluid with viscosity but
zero heat flux (adiabatic but irreversible evolution) whose equilibrium state variables
satisfy the equations of state of: (a) ultra-relativistic ideal gas, (b) non-relativistic
ideal gas, (c) a mixture of (a) and (b). Einstein's field equations reduce to a quadrature
that is integrable in terms of elementary functions (cases (a) and (c)) and elliptic
integrals (case (b)). Necessary and sufficient conditions are provided for the viscous
dissipative stress and equilibrium variables to be consistent with the theoretical
framework of Extended Irreversible Thermodynamics and Kinetic Theory of the
Maxwell-Boltzmann and radiative gases. Energy and regularity conditions are discussed. We
prove that a smooth matching can be performed along a spherical boundary with a FLRW
cosmology or with a Vaidya exterior solution. Possible applications are briefly outlined.} 

\section{Introduction}

The relativistic ideal gas$^{[1]}$ is the main conceptual basis for the study of matter 
sources in cosmology$^{[2,3,4]}$, either in early universe models or in classical
conditions. The various stages of the thermal history of the univese can be described as
the evolution of mixtures of various types of ideal gases, originally very relativistic and
coupled in near thermal equilibrium, later becoming non-relativistic and decoupling. A
gross simplification of this process would group cosmic matter in two main components:
``radiation'' (an ideal gas of massless particles) and ``matter'' (a non-relativistic ideal
gas). Early universe conditions are characterized by ``radiation'' as the dominant source,
while the actual universe is thought of as ``matter''dominated, with ``radiation'' being
present in the form of relics of photon, neutrino and other particles' gases at a very low
equilibrium temperature.

In order to examine this near equilibrium thermal evolution of ideal gas mixtures, 
conventional wisdom among cosmologists has tended so far to use mostly FLRW metrics and
perturbations of them$^{[5]}$, usually assuming a perfect fluid momentum-energy tensor,
with state variables either obtained from equilibrium Kinetic Theory
distributions$^{[2,3,4]}$, or satisfying the so-called ``gamma law'' (a linear relation
between equilibrium pressure and matter-energy density: $p=(\gamma-1)\rho$). From the
perspective of a gamma law, the transition from radiation to matter dominance is loosely
ascribed to the change of the proportionality factor $\gamma$ from $\gamma=4/3$ (radiation)
to $\gamma=1$ (dust). The latter is widely used as a model of present day cosmological
matter, since the pressure of non-relativistic matter is truly negligible.

However, these conventional descriptions of cosmological matter sources have the  following
important weak points: (1) The gamma law obscures the fact that the actual matter model
being considered is that of ideal gas mixtures. Instead, the gamma law equation of state
with constant $\gamma$ is merely a mathematical construct (with the exception of $\gamma=4/3
$). (2) This reasoning applies to dust ($\gamma=1$): the real physical model of dominant
matter content in a late universe is a non-relativistic ideal gas. Dust is merely an
approximated description of such a gas at very low temperatures. (3) The perfect fluid
description (either with a gamma law or with Kinetic Theory equilibrium distributions)
necessarily assumes thermal equilibrium (quasi-static reversible processes). Although early
cosmological gas mixtures are supposed to be in near equilibrium conditions, the small
deviations from equilibrium can be extremely important$^{[2]}$ to account for and explain
interactions between particles, cosmological nucleosynthesis and structure formation.
Besides this fact, we should bear in mind that dissipative transport phenomena are (in
general) present in ideal gases, even classical ones at room temperatures $^{[6]}$.

Therefore, bearing in mind the limitations of conventional treatment of cosmological 
matter sources, we propose in this paper to drop the gamma law, together with perfect fluid
and dust constructs, and to return to the original physically sound matter models: ideal
gases allowing for dissipative fluxes. This requires one to find exact solutions with
``imperfect'' fluid sources in which equilibrium variables satisfy ideal gas equations of
state, and where non-equilibrium variables are consistent with a positive entropy
production law and causal transport equations of irreversible thermodynamics$^{[1,7,8,9]}$. 

A recent paper$^{[10]}$ illustrates a possible strategy to find exact solutions with an 
ideal gas fluid model as matter source. The idea is simple: (1) consider the metric ansatz
of an exact solution usually associated with a dust source, (2) replace the dust source
with an imperfect fluid, (3) impose the equation of state of an ideal gas on the
equilibrium state variables, and finally: (4) verify that non-equilibrium variables are
compatible with causal thermodynamics of irreversible processes. This procedure was
succesfully applied to the spherically symmetric Lem\^aitre-Tolman-Bondi (LTB) metric
ansatz$^{[11,12]}$ and to the equation of state of a non-relativistic monatomic ideal gas.
The resulting models presented two important limitations: (a) shear viscosity is the only
dissipative agent, bulk viscosity and a heat flux are zero (the latter necessarily vanishes
for LTB metrics), and (b) the presence of non-zero pressure gradients with zero
4-acceleration. The lack of bulk viscosity is not a problem (it is negligible for a either
a non-relativistic or an ultra-relativistic ideal gas$^{[7,8,9]}$), but the lack of heat
flux (with non-zero shear viscosity) makes it necessary to assume that the fluid model is
limited to a gas evolving along adiabatic but irreversible processes. Fortunately, the
above mentioned limitations were compensated by the obtention of exact inhomogeneous
solutions with thermodynamical consistency: the shear viscous pressure satisfies the most
general transport equation$ ^{[9,13]}$ (when shear viscosity is the only dissipative flux)
with phenomenological coefficients consistent with those of a Maxwell-Boltzmann gas in the
non-relativistic limit.

In the present paper we attempt to generalize the above mentioned strategy, to a more 
general class of metrics and to encompass other ideal gas equations of state. Among the
most general metric ansatz for a non-rotating dust source we have the class of metrics
associated with the famous Szekeres solutions$^{[11,12,14]}$ (of class I and II), Petrov
type D metrics admitting (in general) no isometries. In particular, the metric of LTB dust
solutions$^{[11,12]}$ mentioned above is the spherically symmetric subcase of Szekeres
class I dust solution. Both classes of Szekeres dust solutions were generalized by
Szafron$^{[11,12,15]}$ to allow for a perfect fluid source characterized by a geodesic (but
expanding and shearing) 4-velocity. This perfect fluid generalization has been called the
``Szekeres-Szafron'' (class I and II) of solutions$^{[11]}$. We shall consider the general
metric ansatz of the class I Szekeres-Szafron solutions, and proceed as mentioned above by
replacing their usual dust (or perfect fluid) source with a non-perfect fluid where the
equilibrium state variables satisfy a generic equation of state comprising the following
cases: (1) an ultra-relativistic ideal gas, (2) a non-relativistic monatomic ideal gas and
(3) a mixture of (1) and (2). Again, the obtained solutions present the limitations
mentioned previously (lack of heat flux, with shear viscosity being the only dissipative
agent). However, the lack of heat flux can be justified for the case (1) and (3) (in the
relativistic limit) by the fact that, at high temperatures, shear viscosity is the dominant
dissipative agent$^{[16]}$. The models are mathematically simple and the conditions for a
consistent thermodynamical intepretation are very similar to those already provided for the
LTB case with a monatomic ideal gas$^{[10]}$. We believe these models are a convenient and
physically consistent improvement on existing matter models based on dust and/or perfect
fluid (with gamma law or not). Finally, the models are ideal theoretical tools to explore
the rich phenomenology associated to the presence of inhomogeneous material sources out of
thermodynamical equilibrium. \\

\section{The relativistic ideal gas.}

A one component relativistic ideal gas is characterized by the equation of state$^{[1]}$

$$
\rho=mc^2n\Gamma(\beta)-nkT,\qquad p=nkT \eqno(1) $$
\[
\Gamma(\beta) \equiv{\frac{K_3(\beta)}{{K_2(\beta)}}},\qquad \beta\equiv{ \frac{mc^2}
{{kT}}}
\]

\n where $\rho,n,T,p$ are matter-energy and particle number densities, absolute temperature 
and equilibrium pressure, $m$ is the particles' mass, $k$ is Boltzmann's constant and $K_2,
K_3$ are second and third order modified Bessel functions of the second kind. The
generalization to an N-component mixture of relativistic ideal gases is straightforward 

$$
\rho=\sum_{A=1}^N{\rho_A},\qquad
\rho_A=m_Ac^2n_A\Gamma(\beta_A)-n_AkT_A,\qquad \beta_A\equiv{\frac{m_Ac^2}{{ kT_A}}} 
\eqno(2a)
$$

$$
p=\sum_{A=1}^N{p_A},\qquad p_A=n_AkT_A \eqno(2b) $$

\n
where an important particular case occurs if there is local thermal equilibrium among the
components, so that $T_A=T$ for all $A$. 

Despite the difficulties in obtaining exact solutions of Einstein's equations complying 
with (1), this equation of state simplifies in the two extremes of the temperature and
energy spectra: the ultra-relativistic (UR) and non-relativistic (NR) regimes,
characterized by $\beta\ll 1$ and $\beta\gg 1$ respectively. Considering the behavior of  
$\Gamma(\beta)$ for these cases, we have

$$
\beta\ll 1,\qquad \Gamma\approx{\frac{4}{\beta}}+{\frac{\beta}{2}}+ \hbox{O}
 (\beta^3),\qquad \rho\approx 3nkT\eqno(3a) $$

$$
\beta\gg1,\qquad \Gamma\approx 1+{\frac{5}{{2\beta}}}+\hbox{O} (\beta^{-2}),\qquad 
\rho\approx mc^2n+{\frac{3}{2}}nkT \eqno(3b) $$

\n
Another interesting limiting case is that of a binary mixture ($A=1,2$) where one 
component is ultra-relativistic and the other non-relativistic (a ``mater-radiation''
mixture), with the pressure of the the UR gas much larger than that of the NR one, but
comparable to the rest mass of the latter. This is obtained by assuming

\[
\beta_1\ll1,\qquad \rho_1\approx 3n_1kT_1 \]
\[
\beta_2\gg1,\qquad \rho_2 \approx m_2c^2n_2+ {\frac{3}{2}}n_2kT_2 \]
\n together with
\[
n_2kT_2\ll n_1kT_1\approx m_2c^2n_2
\]
\n
so that
$$
\rho\approx m_2c^2n_2+3n_1kT_1,\qquad p\approx n_1kT_1 \eqno(3c) $$

\n
In particular, if there exists local thermal equilibrium between the two components  (a
very frequent assumption), we have $T_1=T_2=T$ and so the applicability of this
approximation strongly depends on the ratio of particle number densities $n_1,\,n_2$. For
photons and non-relativistic baryons the ratio $n_2/n_1\approx 10^{-9}$ is a small
number, hence for the temperature range $10^{3}\hbox{K} < T < 10^{5}\hbox{K}$,
characteristic of the radiative era, we have $n_2kT\ll n_1kT$ together with $n_1kT\approx
mc^2n_2$, and so this approximation to a matter-radiation mixture is reasonable$^{[3,16]}$.
If there is no local thermal equilibrium (for example, after decoupling), then $ T_1\ne
T_2$,  but we can still justify (3c) as long as
$n_2kT_2\ll n_1kT_1$ holds.

We shall describe the three cases (3) through a generic equation of state characterized  by
the ansatz

$$
\rho=mc^2n+{\frac{nkT}{{\gamma-1}}},\qquad p=nkT \eqno(4) $$
\n
where $\gamma$ is a positive constant, coinciding in the ultra and non-relativistic  limits
of (1) with the ratio of heat capacities at constant volume and pressure, which in these
limits takes the constant asymptotic values $4/3,5/3$. With this ansatz we can recover
(3a-c) through the following particular cases

$$
\hbox{Ultra-relativistic ideal gas}:\qquad m=0,\quad\gamma={\frac{4}{3}}  \eqno(5a)
$$
$$
\hbox{Matter-radiation mixture}:\qquad m>0,\quad\gamma={\frac{4}{3}}  \eqno(5b)
$$
$$
\hbox{Non-relativistic monatomic ideal gas}:\qquad m>0,\quad\gamma={\frac{5}{3}} \eqno(5c)
$$

\n
where, in the case (5b), it is necessary to bear in mind that $n$ multiplying $mc^2$ is 
$n_2$, while that multiplying $kT$ is $n_1$. However, we shall not use these subindices,
unless the distinction between $n_1,n_2$ is not clearly understood from the context of the
discussion. Also, for classical ideal gases, we can generalize (5c) to $\gamma=1+2/q$, for
$q$ a natural number, describing molecules with $q$ degrees of freedom$^{[17]}$ (the case
(5c) corresponds to $q=3$, a monatomic gas). We show in the following sections that
Einstein's field equations can be integrated for Szekeres-Szafron metrics of class I whose
source is a fluid matter tensor complying with the generic equation of state (4). 

\section{The Szekeres-Szafron metrics.}

The class I of perfect fluid Szekeres-Szafron solutions is characterized by the following 
metric$^{[11,12,14,15]}$

$$
ds^2=-c^2dt^2+(\nu
h)^2\left(Y^{\prime}\right)^2dz^2+Y^2\left[dx^2+dy^2\right]\eqno(6a) $$
$$
Y={\frac{\Phi}{\nu}}\eqno(6b)
$$

\n where $\Phi=\Phi(t,z)$, $h=h(z)$, $Y^{\prime}=Y_{,z}$ and the function  $ \nu=\nu(x,y,z)$
is given by

$$
\nu=A(z)(x^2+y^2)+B_1(z)x+B_2(z)y+C(z) \eqno(6c) $$

\n with $A,B_1,B_2,C$ being arbitrary functions. This metric admits, in general, no 
isometries, but contains as particular cases spherically, plane and pseudo-spherically
symmetric solutions (3-dimensional isometry groups acting on 2-dimensional orbits), as well
as FLRW spacetimes with positive, negative or zero curvature of spacelike slices$^{[11]}$. 

We shall consider as matter source for (6) a viscous fluid characterized by the 
stress-energy tensor

$$
T^{ab}=\rho \,u^au^b+p\,h^{ab}+\pi^{ab}\eqno(7) $$
\[
h^{ab}=g^{ab}+c^{-2}u^au^b,\qquad \pi_{ab}u^b=0,\qquad \pi^a\,_a=0 \]

\n associated with the balance and conservation laws 

$$T^{ab}\,_{;b}=0 \eqno(8a)$$
$$(nu^a)_{;a}=0 \eqno(8b)$$
$$(nsu^a)_{;a}\geq 0 \eqno(8c)$$

\n where $s$ is the entropy per particle. 

In the comoving representation ($u^a=c\delta^a\,_t $) for (6)-(7), the 4-acceleration 
vanishes ($\dot u_a=u_{a;b}u^b=0$) and the remaining non-zero kinematic invariants are the
expansion scalar: $\Theta\equiv u^a\,_{;a}$ and shear tensor: $\sigma_{ab}\equiv
u_{(a;b)}-(1/3)\Theta\,h_{ab}$. These invariants, together with the shear viscous pressure
(a traceless symmetric tensor like $\sigma^a\,_b$), are given by 

$$
\Theta={\frac {\dot Y^{\prime}}{Y^{\prime}}}+{\frac {2\dot Y}{Y}}\eqno(9) $$

$$
\sigma^a\,_b={\bf {\hbox{diag}}}\left[
0,-2\sigma,\sigma,\sigma\right],\qquad \sigma\equiv {\frac {1}{3}}\left({ \frac {\dot
Y}{Y}}-{\frac {\dot Y^{\prime}}{Y^{\prime}}} \right)\eqno(10)   $$
$$\pi^a\,_b={\bf {\hbox{diag}}}\left[ 0,-2P,P,P\right]\eqno(11) $$\\

\n where $\dot Y=u^aY_{,a}=Y_{,t}=\Phi_{,t}/\nu$ and $P=P(t,x,y,z)$ is determined by
the field equations.

The integration of the conservation equation (8b) for (6) leads to 

$$n=\frac{N}{Y^2Y^{\prime}} \eqno(12)$$

\n where $N=N(x,y,z)$ is an arbitrary function (the conserved particle number 
distribution). For the mixture in (5b), we have assumed the two components to have
independently conserved particle number densities and to be characterized by the same
4-velocity, hence: $N=N_1+N_2$ in this case. Einstein field equations for (6) and (7) 
yield 

$$
\kappa\rho=\frac{[Y(\dot{Y}^2+Kc^2)]^{\prime}}{Y^2Y^{\prime}}\eqno(13a) $$
$$
\kappa p=-\frac{[Y(\dot{Y}^2+Kc^2)+2\ddot{Y}Y^2]^{\prime}}{3Y^2Y^{\prime}} \eqno(13b)
$$
$$
\kappa P=\frac{Y}{6Y^{\prime}}\left[\frac{\dot{Y}^2+Kc^2+2\ddot{Y}Y}{Y^2} \right]
^{^{\prime}}\eqno(13c)
$$

\n where $\kappa=8\pi G/c^2$ and the function $K=K(x,y,z)$ is given by 

$$
K=\frac{4h^2\left(AC-B_1^2-B_2^2\right)-1}{h^2\nu^2}\eqno(14) $$

\n This function$^{[11,15]}$ is related to the curvature of the hypersurfaces $t=\hbox
{const}$, everywhere orthogonal to $u^a$. The energy and momentum balance (8a) for (7) is
given with the help of (4) by the following equations

$$
\dot p +\gamma p\Theta+6P\sigma=0 \eqno(15a) $$
$$
6P\frac{Y^{\prime}}{Y}+\left(p+2P\right)^{\prime}=0 \eqno(15b) $$
$$
\left(\frac{\nu^{\prime}}{\nu}\right)_{,x}\left(p+P\right)_{,y} -\left(\frac{
\nu^{\prime}}{\nu}\right)_{,y}\left(p+P\right)_{,x}=0 \eqno(15c) $$

\n where $\nu(x,y,z)$ is given by (6b). We have written equations (13) in terms of $Y$ 
rather than in terms of $\Phi$ and $\nu$, thus obscuring the dependence of the metric on
the coordinates $(x,y)$. However, equation (15c) places a strong restriction on the way in
which initial value functions and state variables may dependend on these coordinates. This
is illustrated in the following section.

\section{Integration of the field equations.} 

Considering the field equations for (6) and (7) satisfying (4), we look first at the 
possibility of a reversible evolution along all fluid worldlines, associated with a perfect
fluid tensor ($\pi^a\,_b=0$ in (7)), and then examine the irreversible case with shear
viscosity. 

\subsection{Perfect fluid case.}

Szekeres-Szafron perfect fluid solutions of class I follow by setting $P=0$ in (13c), 
which together with (13b), yields 

$$
\dot{Y}^2+Kc^2+2\ddot{Y}Y=-\kappa\,p(t)Y^2, \eqno(16) $$

\n where $p(t)$ is the equilibrium pressure, an arbitrary time-dependent function that 
must be specified in order to have determined solutions. Since $\rho$ obtained from (16)
and (13a) depends also on the spatial coordinates, it is an obvious (and a known)
fact$^{[11,15]}$ that these perfect fluid solutions are incompatible with the so-called
``barotropic equations of state'' of the form $p=p(\rho)$. However, (4) is not barotropic,
but a two-parameter equation of state expressible as $\rho=\rho(n,T)$ and $p=p(n,T) $, and
so it is not obvious whether these solutions admit this equation of state or not. It is
also known$^{[18,19]}$ that the class I of Szekeres-Szafron solutions, whose source is a
one-component perfect fluid, is compatible with a two-parameter equation of state and with
the integrability of the equilibrium Gibbs equation only if it admits an isometry group
acting on orbits of two or more dimensions. Therefore, possible perfect fluid
Szekere-Szafron solutions of class I compatible with (4) must necessarily present this  type
of symmetry. The case of a fluid mixture is not subjected to this restriction, but will  not
be discussed. 

Assuming the perfect fluid source ($P=0$) of a Szekeres-Szafron metric to be compatible 
with the integrability of the equilibrium Gibbs equation, equations (15a) and (16) imply
$\Theta=\Theta(t)$, but then (8b), (rewritten as: $\dot{n}+n\Theta=0$), yields $n=n(t)$.
From (4) and (12), we have $ \rho=\rho(t)$ and $T=T(t)$, while (9) implies in this case $
\Phi(t,z)=\Phi_1(t)\Phi_2(z)$, all of which means that the metric reduces to that of a FLRW
spacetime. Therefore, we conclude that a Szekeres-Szafron fluid solution, complying with
(4) and not reducible to a FLRW metric, necessarily requires a momentum-energy tensor with
non-vanishing anisotropic pressure (or shear viscosity). In other words: {\it reversibility
along all fluid worldlines is only possible for the FLRW subcase of (6)} (see section
(9.3)).
\\

\subsection{The viscous fluid case.}

If $P\ne 0$, the field equations (12), (13a) and (13b), subjected to the equation of  state
(4), still provide a closed system and, consequently, Einstein's field equations can be
solved independently of (13c). In this case, equation (12) plus the field equations (13a)
and (13b) subjected to the equation of state (4), yield the constraint 

$$
\left[(3\gamma-2)Y(\dot{Y}^2+Kc^2)+2\ddot{Y}Y^2\right]{}^{\prime}-3\kappa\, (\gamma-1)
\,mc^2N=0 \eqno(17a)
$$
\n which integrates to
$$
\left(3\gamma-2\right)\left(\dot Y^2+Kc^2\right)Y+2\ddot Y Y^2 -3\kappa\left(\gamma-
1\right)M= f(t,x,y),\eqno(17b) $$

\n where $f(t,x,y)$ and $M(x,y,z)=\int{Ndz}$ are arbitrary functions. However, these 
functions are not totaly unrestricted and, as shown in section 9.2, a consistent dust limit
associated with $p=P=0 $ requires $f=0$. Under this assumption the specific functional
dependence of $M$ on $(x,y)$ is given by

$$
M=\frac{\tilde{M}(z)}{\nu^3}. \eqno(18)
$$
This follows from consistency of (17) with (6b) and (14) (once we have set $ f=0$),  and
agrees with (15c). Integrating (17b) again, for $f=0$, we get 

$$
\left(\dot Y^2+Kc^2\right)Y=\kappa M+\frac{J}{Y^{3(\gamma-1)}} \eqno(19) $$

\n where $J=J(x,y,z)$ is an arbitrary integration function. In order to infere the form  of
this function, we substitute $\dot Y^2+Kc^2$ from (19) into (13a) and (using (12) and (18))
compare it with $\rho$ given by (4), resulting in

$$
\kappa\int{\frac{NkT}{\gamma-1}\,\d z}=\frac{J}{Y^{3(\gamma-1)}}\eqno(20a) $$

\n Therefore, by analogy with (18), we can define $J$ in terms of an initial value for  the
function at the left hand side of (20a). This leads to 

$$
J=\kappa U_0Y_0^{3(\gamma-1)},\qquad U_0=\frac{\tilde U_0(z)}{\nu^3}=\int{ \frac{NkT_0}  
{\gamma-1}\,dz} \eqno(20b)
$$

\n so that
\[
U=\int{\frac{NkT}{\gamma-1}\,dz}=\int{\frac{nkT\,Y^2Y^{\prime}}{\gamma-1}\,dz }= 
U_0\left( \frac{Y_0}{Y}\right)^{3(\gamma-1)} \]

\n where, as with (18), the restriction on the dependence of $U_0$ (and thus, $T_0$)  on
$(x,y)$, follows from (6b), (14) and (15c). From now onwards, a subindex``$0$'' will denote
the ``initial value'' of the function it is attached to. That is, the function evaluated
along an arbitrary spacelike hypersurface orthogonal to $u^a$ and marked by the initial
time $ t=t_0$. Using $n_0Y_0^2Y^{\prime}_0=N$ from (12), we can rewrite (18), (19) and
(20b) as

$$
\dot Y^2=\frac{\kappa}{Y}\left[M+U_0\left( \frac{Y_0}{Y}\right)^{3( \gamma-1)}\right]-
Kc^2 \eqno(21a)
$$

$$
M=mc^2\int{n_0Y_0^2Y^{\prime}_0\,dz}\qquad U_0=\int{\frac{ n_0kT_0\,Y_0^2Y^{\prime}_0}
{\gamma-1}}\,dz\eqno(21b) $$

\n in terms of initial value functions $n_0$, $T_0$, $Y_0$ and $K$, with clear physical 
and geometric meaning. However, $n_0$, $T_0$ are not completely general functions since
they must satisfy the restrictions given by (18) and (20b), though $\tilde M,\,\tilde U_0$
are totaly arbitrary and the function $\nu$ contains five extra arbitrary functions of $z$,
so these restrictions basically affect the dependence of $n_0$ and $T_0$ on the coordinates
$(x,y)$. Equation (21) is a generalized Friedmann equation (or evolution equation) for
fluid layers labelled by the comoving coordinates. Again, for the case (5b), the function
$n_0$ (like $n$ and $N$) multiplying $ mc^2$ and $kT_0$ correspond to a separate mixture
component. 

\section{The state variables.}

Before proceeding to integrate (21), it is useful to derive semi-determined forms of the 
state variables and kinematic invariants by inserting (21) into (12) and (13). This yields

$$
n=n_0\left(\frac{Y_0}{Y}\right)^3\frac{Y^{\prime}_0/Y_0}{Y^{\prime}/Y}  \eqno(22)
$$

$$
T=T_0\left(\frac{Y_0}{Y}\right)^{3(\gamma-1)}\Psi\eqno(23) $$

$$
\rho=\left[mc^2n_0+ \frac{n_0kT_0}{\gamma-1}\left(\frac{Y_0}{Y} \right)^{3(\gamma-1)}\,
\Psi\right ] \left(\frac{Y_0}{Y}\right)^3\frac{ Y^{\prime}_0/Y_0}{Y^{\prime}/Y}\eqno(24) $$

$$
p=n_0kT_0\left(\frac{Y_0}{Y} \right)^{3\gamma}\frac{Y^{\prime}_0/Y_0}{
Y^{\prime}/Y}\Psi\eqno(25)
$$

$$
P=\frac{1}{2}n_0kT_0\left(\frac{Y_0}{Y} \right)^{3\gamma}\frac{ Y^{\prime}_0/Y_0}
{Y^{\prime}/Y}\Omega \eqno(26) $$

$$
6\sigma\frac{\dot Y}{Y}\,\frac{Y^{\prime}}{Y}=-\kappa\left[mc^2n\left(1- \frac{3M}
{M^{\prime}}\frac{Y^{\prime}}{Y}\right) +\frac{2P} {\gamma-1}\right]
\frac{Y^{\prime}}{Y}+\frac{Kc^2}{Y^2}\left(\frac{K^{\prime}}{K} -\frac{
2Y^{\prime}}{Y}\right) \eqno(27)
$$
\[
{\frac{{2\dot Y} }{Y}}{\frac{{Y^{\prime}} }{Y}}\Theta =\kappa \left\{ { mc^2n_0} 
\right.\left( {{\frac{{Y_0} }{Y}}} \right)^3\left[ {1+{\frac{{3M} }{
{M^{\prime}}}}{\frac{{Y^{\prime}} }{Y}}} \right]+{\frac{{n_0kT_0} }{{\gamma -1}}}\left(
{{\frac{{Y_0} }{Y}}} \right)^{3\gamma } \]
$$
\left. {\left[ {1+{\frac{{3U_0} }{{U^{\prime}_0}}}\left( {(\gamma -1){\frac{{ Y^{\prime}_0}
}{{Y_0}}}+(2-\gamma ){\frac{{Y^{\prime}} }{Y}}} \right)} \right]}
\right\}{\frac{{Y^{\prime}_0} }{{Y_0}}}-{\frac{{Kc^2} }{{Y^2}}} \left(
{{\frac{{K^{\prime}} }{K}}+{\frac{{4Y^{\prime}} }{Y}}} \right) \eqno(28)
$$

\n where
$$
\Psi\equiv 1+3(\gamma-1)\,\frac{U_0}{U^{\prime}_0}\left(\frac{Y^{\prime}_0}{
Y_0}-\frac{Y^{\prime}}{Y} \right) \eqno(29a) $$
$$
\Omega\equiv 1+\frac{U_0}{U^{\prime}_0}\left[3(\gamma-1)\frac{Y^{\prime}_0}{
Y_0}-3\gamma\frac{Y^{\prime}}{Y} \right] \eqno(29b) $$

\n The new solutions, characterized by (21)-(29), become fully determined once  $Y$,
$Y^{\prime}$ are found by integrating (21) subjected to initial conditions given by the
initial value functions $n_0$, $T_0$, $Y_0$ and $K$. Because of (6b), (6c), (18) and (20b),
the initial value functions are really the $z$ dependendent functions
$A,\,B_1,\,B_2,\,C,\,h$ and $\tilde M,\,\tilde U_0$. Notice that it is always possible to
eliminate any one of these funcions by rescaling the $z$ coordinate. 

\section{General integral and specific solutions.} 

Equation (21) can be transformed into the following adimensional quadrature 

$$
\frac{c}{Y_0}\,(t-t_0)=\pm\int_1^y{\frac{\bar y^{(3\gamma-2)/2}\,d\bar y}{ \left[-K \bar
y^{3\gamma-2}+\mu \bar y^{3\gamma-3}+\omega \right]^{1/2}}}  \eqno(30)
$$

\n where:
\[
y\equiv \frac{Y}{Y_0}=\frac{\Phi}{\Phi_0},\qquad \mu\equiv \frac{\kappa M}{
c^2Y_0}=\frac{\kappa\tilde M}{c^2\nu^2\Phi_0},\qquad \omega\equiv \frac{ \kappa
U_0}{c^2Y_0}=\frac{\kappa\tilde U_0}{c^2\nu^2\Phi_0} \]

\n so that (30) can also be though of as the integral yielding $\Phi$ (which, from (6a) 
yields $Y$). We integrate (30) explicitly for the three cases of interest mentioned in
equations (5). From the results of these integrals it is possible to obtain the gradients
$Y^{\prime}$ and to evaluate explicitly (22)-(29).

\subsection{Ultra-relativistic ideal gas.} 

Setting $\gamma=4/3$ and $m=0$ (or, equivalently $\mu=0$) in (30), we obtain the  following
solutions\\

\n{\bf{Case $K=0$}}

$$
Y=Y_0\left[1\pm\frac{2c\sqrt{\omega}}{Y_0}(t-t_0) \right]^{1/2} \eqno(31a) $$

\n{\bf{Case $K\ne 0$}}

$$
\frac{c}{Y_0}(t-t_0)=\pm K^{-1}\left[\sqrt{\omega-K}-\sqrt{\omega-Ky^2} \right]\eqno(31b)
$$

\subsection{Matter-radiation mixture.}

This case is similar to the previous one: $\gamma=4/3$ but $\mu>0$ in (30) (so, $M>0$).  As
mentioned previously, for this case the functions $n_0$, $n$ or $N$ multiplying $mc^2$ or
$kT_0$ correspond to each separate component of the mixture.\\

\n{\bf{Case $K=0$.}}

$$
\frac{3}{2}\sqrt{\mu}\frac{c}{Y_0}(t-t_0)=\sqrt{y+\delta} \left(y-2\delta \right)- 
\sqrt{1+\delta}\left(1-2\delta\right) \eqno(32a) $$
\n where
\[
\delta\equiv\frac{\omega}{\mu}=\frac{U_0}{M} \]

\n{\bf{Case $K\ne 0$.}}

$$
\frac{c}{Y_0}(t-t_0)=\frac{\pm 1}{K}(\sqrt{\mu+\omega-K}-\sqrt{-Ky^2 +\mu y+\omega})+ 
\Lambda\eqno(32b)
$$

\[
\Lambda= \ln\left[\frac{2\sqrt{K^2y^2-K\mu y-K\omega}-2Ky+\mu}{2\sqrt{ K^2-K\mu-K\omega}-
2K+\mu}\right],\qquad K<0 \]
\[
\Lambda=\arcsin\frac{\mu-2K}{\sqrt{-\Delta}}-\arcsin\frac{\mu-2Ky}{\sqrt {-\Delta}}, \qquad
K>0
\]
\[
\Delta=\mu^2-4K\omega
\]

\subsection{Non-relativistic monatomic ideal gas.} 

This case follows by setting: $\gamma=5/3$ and $\mu>0$ in (30). The corresponding 
integrals are combinations of elliptic integrals and algebraic functions. They have been
obtained with the help of standard text books dealing with these integrals$^{[20]}$\\

\n{\bf{Case $K=0$.}}

$$
{\frac{3 }{2}}\sqrt {\mu}{\frac{c}{Y_0}}\left( {t-t_0} \right)=\sqrt {y\left(  {y^2+\delta
} \right)} -\sqrt {1+\delta }+{\frac{{\delta ^{3/ 4}} }{ 2}}\left( {\hbox{F}}-{\hbox{F}_0}
\right) \eqno(33a) $$

\n where
\[
\delta\equiv \frac{\omega}{\mu}=\frac{U_0}{M} \]

\n and ${\hbox{F}}$ is the elliptic integral of the first kind with modulus $ 1/\sqrt{2}
$ and argument $\varphi$ given by: $\cos\varphi=(y-\sqrt{\delta} )/(y+\sqrt{\delta}) $.
(see equation 241.00 of reference [20]). \\ 

\n{\bf{Case $K\ne 0$.}}

\n This case follows from equations 260.03 and 341.05 of reference [20].  It requires
re-writing (30) as

\[
\frac{c}{Y_0}(t-t_0)=\pm \int_1^y{\frac{\bar y^2 d\bar y}{\left[\bar y\left(-K\bar 
y^3+\mu\bar y^2+\omega\right)\right]^{1/2}}} \]
$$
=\pm\int_1^y{\frac{\bar y^2 d\bar y}{\left\{-K\bar y\left(\bar y-b\right) \left[\left(\bar
y-b_1\right)^2+a_1^2\right]\right\}^{1/2}}} \eqno(33b)
$$

\n which integrates to

$$
\frac{c}{Y_0}(t-t_0)=\pm \frac{bU}{\left(U+V\right)^2}\,g\,\left[\Delta( \phi,\epsilon)-
\Delta(\phi_0,\epsilon)\right] \eqno(33c) $$

\n where
\[
\Delta(\phi,\xi)\equiv\hbox{F}(\phi,\epsilon)-\frac{4V}{V-U} \left[\hbox{R}
^{(1)}(\phi,\epsilon)- \frac{V}{V-U}\hbox{R}^{(2)}(\phi,\epsilon)\right] \]

\[
\hbox{R}^{(m)}(\phi,\epsilon)\equiv \int_0^\phi{\frac{d\vartheta}{
\left(1+\zeta\cos\vartheta\right)^m \sqrt{1-\epsilon^2 \sin\vartheta}}} \]

\n and $\hbox{F}(\phi,\epsilon)$ is the elliptic integral of the first kind with argument 
$\phi$ and modulus $\epsilon$. The latter parameters plus the remaining quantities in (33b)
and (33c) are 

\[
b=\frac{\alpha^2+4\mu^2+2\mu\alpha}{6K\alpha},\quad b_1=-\frac{ \alpha^2+4\mu^2-4
\mu\alpha} {12K\alpha},\quad a_1=\frac{\sqrt{3} (4\mu^2-\alpha^2)}{12K\alpha}
\]

\[
U^2=\frac{\alpha^4-2\mu\alpha^3-8\mu^3\alpha+16\mu^4}{36\alpha^2},\qquad V^2=
\frac{\alpha^4 +4\mu^2\alpha^2+16\mu^4}{12\alpha^2} \]

\[
\cos\phi=\frac{6K\alpha\left[(2\mu-\alpha)- \sqrt{3(\alpha^2-2\mu\alpha+4 \mu^2)}
\right]y-(2\mu-\alpha)(\alpha^2+2\mu\alpha+4\mu^2)} {6K\alpha\left[(2\mu-\alpha)+
\sqrt{3(\alpha^2-2\mu\alpha+4\mu^2)} \right]y-(2\mu-\alpha)(\alpha^2+2\mu\alpha+4\mu^2)} \]

\[
g=\frac{1}{\sqrt{UV}},\qquad \epsilon^2=\frac{(U+V)^2-b^2}{4UV},\qquad\zeta= \frac{V+U}{V-U}
\]

\[
\alpha^3=108\omega+8\mu^3+12\sqrt{3}K\left[\omega(27\omega +4\mu^3)\right]^{1/2}
\]

\section{Conditions for thermodynamical consistency.} 

We have obtained a class of exact solutions whose equilibrium variables satisfy the 
equations of state (3)-(5). The entropy per particle, $s$, and the shear viscous pressure,
$\pi_{ab} $, (the only dissipative flux present in the matter tensor) were not used in the
integration of the field equations. It is necessary now to examine the consistency of these
quantities within the framework of a physically acceptable theory of irreversible
thermodynamics. Concretely, this means verifying that $\pi_{ab}$ and $s$ satisfy a suitable
transport equation and the entropy balance law (8c), for physically reasonable
phenomenological coefficients compatible with (3)-(5).

The oldest approach to irreversibility is the so called ``classical'' theory of 
irreversible thermodynamical processes$^{[7]} $, based on the hypothesis of ``Local
Equilibrium''. This implies considering $s$ as depending only on equilibrium variables and
obtained from the integration of the equilibrium Gibbs equation:
$T\hbox{d}s=\hbox{d}(\rho/n)+p\hbox{d}(1/n)$. For: $ n,\,T,\,\rho,\,p$ related by the
equation of state (4), this means 

$$
s^{(e)}=s_0+k\ln\left[\left({\frac{{T}}{{T_0}}}\right)^{1/(\gamma-1)} {\frac{ {n_0}}
{{n}}}\right]=s_0+k \ln\left[ {\frac{{Y^{\prime}/Y}}{{Y^{\prime}_0/Y_0} }}\,
\Psi^{1/(\gamma-1)}\right]\eqno(34) $$

\n where $s_0$ is an initial value function. In order to apply (34) to the case (5b), we 
have assumed a single $T$ for the mixture (local thermal equilibrium between the
components). In this case, $n,n_0$ appearing in (34) can be considered particle number
densities of the ultra-relativistic component.

It is a well known fact that equation (34), together with (8c), yield parabolic transport 
equations$^{[7-9]} $, and so the classical theory is applicable to physical systems whose
microscopic time and length scales (mean free path and mean collision time) are much
smaller than the macroscopic evolution times or characteristic length scales. However,
these conditions are not always applicable to astrophysical and cosmological systems, both
non-relativistic and relativistic. The former (globular clusters, galaxies, dust clouds)
are practically collisionless, or have mean collision times comparable to their evolution
time, the latter (early universe gas mixtures) cannot be compatible with transport
equations violating causality. The alternatives to the classical theory are the theories
generically grouped under the term``Extended (or causal) Irreversible Thermodynamics''
(EIT)$^{[7-9]} $.

EIT assumes that $s$ also depends on the dissipative stresses. For the momentum-energy 
tensor (7), where shear viscosity is the only dissipative stress, EIT associates a
generalized entropy current satisfying (8c) and relating the deviation from equilibrium  due
to viscosity $$ (nsu^a)_{;a}\geq0\Rightarrow \dot s\geq0, \qquad s=s^{(e)}-\alpha
\pi_{ab}\pi^{ab}\eqno(35)
$$
\noindent where $\alpha$ is a phenomenological coefficient and $s^{(e)}$ is given by (34). 
If shear viscosity is the only dissipative stress, the most general available transport
equation in the formalism of EIT is$^{[9,13]}$ $$
\tau \dot\pi_{cd}\,h^c_ah^d_b+\pi_{ab}\left[ 1+{\frac{1}{2}}T\eta\left({ \frac{{\tau}}
{{T\eta}}} \,u^c\right)_{;c}\right]+ 2\eta\,\sigma_{ab}=0 \eqno(36)
$$
\noindent where $\dot\pi_{ab}\equiv\pi_{ab;c}u^c$, and $\tau,\eta$ are the relaxation time 
and the coefficient of shear viscosity, phenomenological quantities whose form depends on
the properties of the fluid. The transport equation of the classical theory is recovered
from (36) by assuming $\tau=0$ . Also, within EIT, a simpler or ``truncated form'' without
the term involving $u^c$ and $T$ in (36) is often suggested$^{[7,8]}$, for no other reason
than to deal with a mathematically simpler transport equation. However, althogh this
truncated form corrects the problems of acausality and unstability plaging the classical
theory, we shall consider the full transport equation (36), since numerical studies with
FLRW spacetimes$^{[21]} $ and other theoretical arguments$^{[22]}$ reveal adverse physical
effects, and so, advice caution in using truncated versions of viscosity transport
equations.

Equations (35) and (36) couple to Einstein's equations through the dependence of state 
variables on the metric functions. However, the fulfilment of these equations cannot be
verified until an explicit form for the phenomenological quantities $\alpha,\tau,\eta$ is
somehow specified. As a referencial value of the forms that one should expect for these
quantities in dealing with one-component ideal gases associated with equations of state
like (3)-(5), consider a relativistic Maxwell-Boltzmann distribution near equilibrium for
which the equation of state (1) holds$^{[7,8]}$. From equations (7.1) and (7.10c) of
reference [8], and bearing in mind that the factor multiplying $\dot\pi_{ab}$ is $\tau$, we
obtain $\eta$ in terms of $ \tau$ as
$$
\eta=\frac{\tau}{2\beta_2}=\frac{\Gamma^2(\beta)}{1+6\Gamma(\beta)/\beta}\ \,p\,\tau 
\eqno(37a)
$$
\n where $\Gamma(\beta)$ is given in equation (1) \footnote{ The quantities $(\eta,\beta_2,
\Gamma,p)$ here are respectively given by $ (\zeta_S,\beta_2,\eta,P)$ in reference [8]}.
Regarding the case (5b), the most convenient referential system for such a gas mixture is
the ``radiative gas''$^{[4,7]}$, a mixture of photons and non-relativistic particles, for
which we have:

$$
\eta=\frac{4}{5}\,p\,\tau\eqno(37b)
$$

\n where $p$ is the pressure of the ultra-relativistic component (photon gas). For both 
cases, (37a) and (37b), the coefficient $\alpha$ is given by $$
\alpha=\frac{\tau}{4\eta\,nT}\eqno(37c)
$$
\n Considering the expansion of $\Gamma(\beta)$ in the ultra-relativistic ($ \beta\ll 1$) 
and non-relativistic ($\beta\gg 1$) limits in (37a), we can provide a compact expression
for $\eta,\alpha$ comprising these limits (cases (5a) and (5c)), as well as the radiative
gas (case (5b)) $$
\eta =b_0\,p\tau ,\quad \quad \alpha ={\frac{1 }{{4b_0\,pnT}}}={\frac{k }{{ 4b_0\,p^2}}} 
,\quad \quad b_0=\left\{ \matrix{2/ 3,\quad \hbox{Ultra-Rel.} \hfill\cr 4/ 5, \quad
\hbox{Mixture} \hfill\cr 1, \qquad \hbox{Non-Rel.}\hfill\cr} \right.\eqno(37d) $$

\n where $p=nkT$ was used in (37d). The relaxation time $\tau$, for all these cases,  is
given by collision integrals (for example, see eqs (3.41) and (3.45a) of [7] for the
non-relativistic case). 

The consistency of the solutions with EIT and Relativistic Kinetic Theory follows from 
verifying to what degree the forms of $T$, $p$, $P$ and $\sigma$ , as given by equations
(22)-(29), are compatible with (35), (36) and the coefficients (37). Substitution of
(21)-(29) into (35), with $s^{(e)}$ and $ \alpha$ given by (34) and (37c) yields
\[
s=s_0+k\ln\left[\left({\frac{{T}}{{T_0}}}\right)^{1/(\gamma-1)} {\frac{{n_0} }{{n}}}
\right]-{\frac{{3k} }{{2b_0}}}\left({\frac{{P}}{{p}}}\right)^2 \]
$$
=s_0+k \ln\left[ {\frac{{Y^{\prime}/Y}}{{Y^{\prime}_0/Y_0}}}\, \Psi^{1/(\gamma-1)}\right]
-{\frac{3k}{{8b_0}}}\left({\frac{{\Omega}}{{\Psi}}} \right)^2 \eqno(38)
$$
\noindent where $s_0(r)$ is an arbitrary initial value of $s$. Regarding (36), if we 
assume the Maxwell-Boltzmann form: $\eta=b_0\,p\,\tau$ given by (37d), with $p$ given by
(23), then this transport equation is satisfied for $\tau$ given by:
$$
\tau=-{\frac{{\Omega\Psi}
}{{4b_0\sigma}}}\left\{\left[\Psi-\frac{9}{8b_0}\,\frac{U_0}{U'_0}\,\frac{Y'}{Y}\right]^2
+b_1\left(\frac{U_0}{U^{\prime}_0}\,\frac{Y^{\prime}}{Y}
\right)^2
\right\}^{-1}\eqno(39)
$$
\[
b_1\equiv \frac{27}{4b_0} \left(\gamma-1-
\frac{3}{16b_0} \right)
\]
While the form of $\eta$ is formally identical to the Maxwell-Boltzmann shear viscosity 
coefficient (37d), and thus needs no justification, the form of $\tau$ in (39) is
acceptable as long as this expression behaves as a relaxation parameter for the system,
namely: it should be positive for all fluid worldlines, and must guarantee that $\dot s$
(computed from (38)) is also positive everywhere. Ideally, of course, $\tau$ should also,
somehow, approach or relate to its corresponding collision integrals from Kinetic Theory of
the Maxwell-Boltzmann relativistic gas. It must also relate to the mean collision time, and
so should increse (decrease) as the fluid expands (collapses). Evaluating $\dot s$ from
(37) and comparing with (39), we find that $\dot s$ and $\tau$ satisfy

$$
(nsu^a)_{;a}=\frac{\pi_{ab}\pi^{ab}}{2\eta T},\quad\Rightarrow\quad \dot s= \frac{3k}
{b_0\tau} \left(\frac{P}{p}\right)^2 =\frac{3k}{4b_0\tau}\left(
\frac{\Omega}{\Psi}\right)^2\eqno(40)
$$

\noindent a relation that also follows directly from (35) and (36). Equation (40) implies: 
$\dot s>0\Leftrightarrow \tau>0$. Therefore, since $b_1>0$ in (39) for the cases of
interest ($b_0=2/3,4/5,1$), equations (39) and (22)-(27) lead to the following necessary
and sufficient conditions for $ T>0,\tau>0,\dot s>0$
$$
\Psi>0\eqno(41a)
$$
$$
\sigma\Omega<0\eqno(41b)
$$
\noindent while the conditions insuring that $\dot s$ decreases for increasing $\tau$ 
($\dot\tau>0 \Leftrightarrow \ddot s<0$) follow from (39)-(40)
$$
\dot \tau>0,\qquad {\frac{\ddot s}{\dot s}}={\frac{{2\sigma}}{{\Psi\Omega}}} \,{\frac{U_0}
  {U^{\prime}_0}}{\frac{Y^{\prime}}{Y}}\left[1+3(\gamma-1){\frac{ U_0}{U^{\prime}_0}}
{\frac{Y^{\prime}_0}{Y_0}}\right]-{\frac{\dot\tau}{{\tau} }}<0\eqno(41c)
$$
\noindent The set (41) provides the necessary and sufficient conditions for a theoretically 
consistent thermodynamical description of the solutions complying with cases (5a-c) of the
generic equation of state (4), within the framework of equations (34)-(40). The
verification of the fulfilment of (41) requires the explicit computation of the gradients
$Y^{\prime}/Y$ from (31)-(33). However, fulfiling this task in the present paper would make
it too long, hence we believe it is more convenient to present here the general theoretical
outline and leave a proper full examination of thermodynamical features for each case (5)
in separate papers$^{[10], [23]}$. In particular, it is worthwhile to mention that
physically reasonable numerical examples satisfying (41) have been found$^{[10],[23]}$ for
the spherical subcase of (6) and the equations of state (5a) and (5b). 

\section{Energy conditions and regularity.} 
Together with the conditions for thermodynamical consistency, the solutions must comply 
with energy and regularity conditions. From (22)-(29) it is evident that $Y/Y_0=0$ marks a
scalar curvature singularity characterized by the blowing up to infinity of all state
variables. This singularity is analogous to a big bang in FLRW cosmologies, and has been
reported in Szekeres-Szafron solutions with dust and perfect fluid sources$^{[24],[25]}$ .
As in these cases, it is not marked (in general) by a constant value of $t$ . However, an
important requirement for the present models is to select an everywhere regular initial
hypersurface, so that $t=t_0$ lies entirely in the future (past) of $Y/Y_0=0$ for expanding
(collapsing) configurations. 

Another scalar curvature singularity, of the so called ``shell crossing'' type, occurs if 
the gradient $Y^{\prime}/Y$ vanishes. Since this gradient appears in the denominator of
$n$, $p$ and $P$, but does not appear in the expression for $T$, if $Y^{\prime}/Y=0$ we
would have $n$, $p$ and $P$ diverging but (in general) finite $T$. This is, obviously, a
totaly unacceptable situation from a physical point of view. Therefore, we must demand the
following regularity condition 

$$
\frac{Y^{\prime}/Y}{Y^{\prime}_0/Y_0}>0\eqno(42) $$

\noindent to hold along the entire evolution range being considered. Notice that the 
possibility of having an opposite relation sign in (42) is ruled out because $n$ would be
negative and, for $p>0$, it would imply $\Psi<0$ and so, $T$ would be negative.

The eigenvalues of the momentum-energy tensor (7) are: $\lambda_0=-\rho$, $ \lambda_1=
p-2P$, $\lambda_2=\lambda_3=p+P$, hence the weak, dominant and strong energy conditions
(WEC, DEC and SEC) are given by 

$$
\rho\ge 0,\qquad \rho+p-2P\ge 0. \qquad \rho+p+P\ge 0, \qquad \hbox{WEC}  \eqno(43a)
$$
$$
\rho\ge 0,\qquad |p-2P| \le \rho, \qquad |p+P| \le \rho, \qquad \hbox{DEC}  \eqno(43b)
$$
$$
\hbox{DEC plus:}\qquad\qquad \rho+3p\ge 0, \qquad\qquad \hbox{SEC} \eqno(43c) $$

\noindent With the help of (4) and (22)-(25), and assuming (42) to hold, we can translate 
(43) in terms of the parameters of the solutions, leading to the following equivalent set
of conditions\\ 

\noindent WEC
$$
mc^2n_0\left(\frac{Y}{Y_0}\right)^{3(\gamma-1)}+\frac{n_0kT_0}{\gamma-1} \Psi\ge 0 
\eqno(44a)
$$
$$
mc^2n_0\left(\frac{Y}{Y_0}\right)^{3(\gamma-1)}+n_0kT_0\left(\frac{\gamma} {\gamma-1}
\Psi-\Omega \right)\ge 0\eqno(44b) $$
$$
mc^2n_0\left(\frac{Y}{Y_0}\right)^{3(\gamma-1)}+\frac{n_0kT_0}{2}\left(\frac {2\gamma} 
{\gamma-1}\Psi+ \Omega \right)\ge 0\eqno(44c) $$

\noindent DEC
$$
n_0kT_0\, |\Psi-\Omega|\, \le\, mc^2n_0\left(\frac{Y}{Y_0} \right)^{3(\gamma-1)}+
\frac{n_0kT_0}{\gamma-1}\Psi \eqno(44d) $$
$$
\frac{3}{2}n_0kT_0\, |2\Psi+\Omega|\, \le\, mc^2n_0\left(\frac{Y}{Y_0} \right)^
{3(\gamma-1)} +\frac{n_0kT_0}{\gamma-1}\Psi \eqno(44e) $$

\noindent SEC
$$
mc^2n_0\left(\frac{Y}{Y_0}\right)^{3(\gamma-1)}+\frac{3\gamma-2}{\gamma-1} n_0kT_0 \Psi\ge
0 \eqno(44f)
$$

\noindent where in the case of the matter-radiation mixture (5b), the initial densities 
$n_0$ multiplying $mc^2$ and $kT_0$ respectively correspond to the non-relativistic and
ultra-relativistic components. Notice that conditions (41a) and (42) are sufficient for the
fulfilment of (44a) and (44f), while the fulfilment of the remaining conditions can be
examined by reducing, with the help of (21b), (29a) and (29b), the terms appearing in (44)
to the following forms

$$
\Psi-\Omega=\frac{3(\gamma-1)U_0}{n_0kT_0Y_0^3}\frac{Y^{\prime}/Y}{ Y^{\prime}_0/Y_0} 
\eqno(45a)
$$
$$
\frac{\gamma}{\gamma-1}\Psi-\Omega=\frac{1}{\gamma-1}+\frac{3(\gamma-1)U_0 }{ n_0kT_0Y_0^3}
\eqno(45b)
$$
$$
2\Psi+\Omega=3\left[1+\frac{3(\gamma-1)^2U_0}{n_0kT_0Y_0^3}\left( 1-\frac{ 3\gamma-2} 
{3(\gamma-1)}\frac{Y^{\prime}/Y}{Y^{\prime}_0/Y_0}\right) \right] \eqno(45c)
$$
$$
\frac{2\gamma}{\gamma-1}\Psi+\Omega=\frac{3\gamma-1}{\gamma-1}\left[1+ \frac{ 
3(\gamma-1)^2U_0} {n_0kT_0Y_0^3}\left( 1-\frac{3\gamma}{3\gamma-1}\frac{
Y^{\prime}/Y}{Y^{\prime}_0/Y_0}\right) \right] \eqno(45d) $$

\noindent
It is evident that the rhs of (45b) is positive for $\gamma>1$, while (42) is  sufficient
for the terms in the rhs of (45a) to be positive, and so (41a) and (42) are sufficient for
the fulfilment of (44b) and (44d). However, nothing can be said about (44c) and (44e) until
the gradient $Y^{\prime}/Y$ is evaluated. The conditions for thermodynamic consistency do
not restrict the sign and magnitude of $\Omega$, as long as (41b) holds, but conditions
(44c) and (44d), through (45c) and (45d), do provide further restrictions on the gradient
of $Y$ controlling how much bigger can $|\Omega|$ be with respect to $\Psi$, or how large
can be $|P|/p$. Therefore, if $|P|/p\gg1$, then the WEC (DEC) might be violated if $P<0$
($P>0$). However, this ratio between $P$ and $p$ is unphysical since the
linear-thermodynamics approach, as that of EIT, demands that $|P|/p\ll 1$ because of the
near-equilibrium conditions on which the theory is based on. Like the fulfilment of
conditions (41), the fulfilment of (44) for each specific case will be examined in separate
papers.

\section{Spherical, dust and FLRW subcases.} 

The solutions derived and classified in the previous sections contain various important 
particular cases, obtained either by imposing extra symmetries on (6), by restricting the
matter sources or both. For the purpose of describing these subcases, it is useful to
transform the spatial coordinates of metric (6) to spherical coordinates. This is achieved
by the transformations: $z=r$, $x=2\tan(\theta/2)\cos(\phi)$, $y=2\tan(\theta/2)
\sin(\phi)$, leading to

$$
ds^2=-c^2dt^2+h^2\nu^2\left(Y^{\prime}\right)^2dr^2+Y^2\left[d\theta^2+\sin^ 2\theta 
d\phi^2\right] \eqno(46)
$$

\[
Y=\frac{\Phi}{\nu},\qquad \nu=4A\sin^2(\theta/2) +2\sin\theta(B_1\cos\phi+B_2\sin\phi)
+C\cos^2(\theta/2) \]
\[
M=\frac{\tilde M(r)}{\nu^3},\qquad U_0=\frac{\tilde U_0(r)}{\nu^3} \]
\vskip 0.5cm
\n where $\Phi=\Phi(t,r)$, $h,A,B_1,B_2,C$ are now functions of $r$ and a prime denotes 
now derivative wrt $r$.

\subsection{Spherical symmetry.}

From (46), it is clear that spherically symmetric subcases of the solutions 
follow by especializing the functions $A,B_1,B_2,C$ to the values 

$$
B_1=B_2=0,\quad A^{\prime}=C^{\prime}=0,\quad 4A=C\eqno(47a) $$

\n leading to
$$
\nu=1, \qquad K=C^2-\frac{1}{h^2} \eqno(47b) $$
$$
M=\tilde M(r),\qquad U_0=\tilde U_0(r) \eqno(47c) $$

\n while $h=h(r)$ remains arbitrary. In fact, the deviation of $A,B_1,B_2,C$ from  the
values in (47a) ``gauges'' the deviation of the geometry from spherical symmetry. The
metric resulting from applying (47) into (46) is known as the
Lem\^aitre-Tolman-Bondi$^{[11,12]}$ metric ansatz, and it is usually associated with dust
solutions. However, this ansatz can also be the metric of exact solutions with a perfect
fluid source (spherical subcase of Szekeres-Szafron class I solutions) or with the type of
viscous source examined in the present paper. Identifying: $\Phi=Y$ and
$h^2=1/(1-k_0f^2(r)) $, leads to the solution examined in reference [10], where only the
equation of state (5c) was considered. Other particular cases characterized by isometry
groups acting on 2-dimensional orbits (plane and pseudo-spherical symmetries) can be
obtained by specializing the functions $A,B_1,B_2,C$ to specific constant values (see [11]).

\subsection{Dust subcases and dust limit.} 

If $n_0>0$ but $T_0=0$ (or equivalently: $M>0$ and $U_0=0$), we have $T=p=P=0 $ with 
$\rho=mc^2n= M^{\prime}/(Y^2Y^{\prime})$, leading to the Szekeres class I dust
solution$^{[11,12,14]}$. If besides these restictions on the state variables we include
spherical symmetry, we obtain the Lem\^aitre-Tolman-Bondi dust solutions$^{[11,12]}$. 

However, the solutions must have a consistent dust limit, by which we shall mean that 
setting $p=0$ in equations (13) and (15) yields exactly the same results characterizing
Szekeres dust solutions or their particular cases with isometries. We prove below that, as
mentioned in section 4.2, a consistent dust limit for the solutions requires $f(t,x,y)=0$
in (17b). First, we re-write (17b) as

\[
Y(\dot Y^2+Kc^2)+2\ddot YY^2= f(t,x,y)+3(\gamma-1)\left[ \kappa M-Y(\dot Y^2+Kc^2)\right]
\]

\n so that substitution into equations (13b) and (13c) yields 

\[
\kappa p=-3(\gamma-1)\frac{\left[\kappa M-Y(\dot Y^2+Kc^2)\right]{}^{\prime} }{Y^2Y^
{\prime}}
\]

\[
\kappa P=\frac{Y}{6Y^{\prime}}\left\{ f\left(\frac{1}{Y^3}\right){}^{\prime}+
3(\gamma-1)\left[\frac{\kappa M-Y(\dot Y^2+Kc^2)}{Y^3}\right]{} ^{\prime}\right\}
\]

\n The condition $p=0$ yields
\[
\dot Y^2=\frac{\kappa M-g(t,x,y)}{Y}-Kc^2 \]

\n which coincides with the Friedmann-like evolution equation of Szekeres dust solutions 
only if $g=g(x,y)$, but then this function can be absorbed into $M$, hence no generality is
lost by setting $g=0$, leading to 

\[
\kappa P=\frac{Y}{6Y^{\prime}}\, f \left(\frac{1}{Y^3}\right){}^{\prime} \]

\n but from (15a), $p=0$ implies $P=0$ (or $\sigma=0$ leading to FLRW subcases, for which 
$P$ also vanishes), therefore we must have $f=0$ for a consistent dust limit.

The solutions are also consistent with the intuitive idea that dust is an asymptotic and 
low temperature limit of an expanding ideal gas. In fact, by looking at the Friedmann-like
equation (21), it is evident that for $m>0$ and $\gamma>1$, this equation tends, for
$Y/Y_0\gg 1$, to its equivalent evolution equation for dust. Similarly, the form of the
state variables (22)-(29) reveals that, in these asymptotic conditions, the equilibrium  and
shear viscous pressure decay much faster than $mc^2n$, the rest mass energy. Therefore, the
solutions behave asymptotically like dust solutions with the same sign for the function $K$.

\subsection{FLRW subcases.}

If together with spherical symmetry, we assume homogeneity by imposing ``homogeneous'' 
initial conditions: $n_0=\bar n_0$ and $T_0=\bar T_0$, where $\bar n_0,\bar T_0$ are
positive constants, we have: 

$$
\bar M=\frac{mc^2\bar n_0\,Y_0^3}{3}, \quad \bar U_0=\frac{\bar n_0 k\bar T_0\,Y_0^3}
{3(\gamma-1)}\eqno(48a)
$$

\n If we also assume $K$ in (47b) to take the form: $K=k_0Y_0^2$, we can re-write (21) as

$$
\dot y^2=\frac{\kappa}{y}\left[\bar\mu+\frac{\bar\omega}{y^{3(\gamma-1)}} \right]- 
k_0c^2, \quad \bar\mu=\frac{\bar M}{Y_0^3}, \quad \bar\omega= \frac{ \bar
U_0}{Y_0^3},\quad y=\frac{Y}{Y_0}\eqno(48b) $$

\n so that its solutions will be of the form $y=f(t-t_0)$, and $Y=\Phi$ is a separable 
function given by $Y=(R/R_0)Y_0$, where $R=R(t)$ and $R_0=R(t_0)$.  From (47b) we have:
$1/h^2=1-k_0Y_0^2$, and (46) becomes a FLRW metric 

$$
ds^2=-c^2dt^2+\left(\frac{R}{R_0}\right)^2\left[dr^2+Y^2_0(r)\left(d\theta ^2+\sin^2 \theta
d\phi^2\right)\right] \eqno(48c) $$

\n where $Y^{\prime}_0=\sqrt{1-k_0Y_0^2}$, so that: $Y_0=(r,\sin r, \hbox{sinh}\,r)$  for
$k_0=(0,1,-1)$, Inserting $Y=(R/R_0)Y_0$ into (22)-(29), we obtain $\sigma=P=0$,
$\Theta=3R_{,t}/R$, together with 

\[
n=\bar n_0\left(\frac{R_0}{R}\right)^3,\quad \rho=\left(\frac{R_0}{R} \right)^3\left[\bar 
M+\bar U_0\left(\frac{R_0}{R}\right)^{3(\gamma-1)} \right]
\]
$$
kT=k\bar T_0\left(\frac{R_0}{R}\right)^{3(\gamma-1)}, \quad p=\bar (\gamma-1)\bar U_0
\left(\frac{R_0}{R}\right)^{3\gamma}\eqno(48d) $$

\n and so, the source of this FLRW spacetime is a perfect fluid ($P=0$) satisfying the 
generic equation of state (4)$^{[26]}$. This particular case is the equilibrium limit of
the solutions, characterized by fluid models of ideal gases following quasi-statical
reversible processes along all fluid worldlines. Although the dust subcase can also be
considered an equilibrium limit, it is a singular type of equilibrium since $T=0$. 

\section{Matching along a spherical boundary.} 

We illustrate in this section the fact that the solutions can be smoothly matched along a 
spherical interface, either to a spherical or FLRW subcases, or to the Vaidya
spacetime$^{[11,12]}$. Since spherical symmetry follows by specializing the functions
$A,B_1,B_2,C$ to the values given by (47), and since these functions determine
$\nu,\,n_0,\,T_0$ and are entirely arbitrary, it is possible to prescribe them so that
spherical symmetry is smoothly reached for a given arbitrary $r=r_{_B}$. This requires the
following boundary conditions associated with (47) on $A,B_1,B_2,C$ 

$$
B_1(r_{_B})=B_2(r_{_B})=0,\quad B^{\prime}_1(r_{_B})=B^{\prime}_2(r_{_B})=0 \eqno(49a)
$$
$$
4A(r_{_B})=C(r_{_B})=\hbox{const.}\quad
A^{\prime}(r_{_B})=C^{\prime}(r_{_B})=0 \eqno(49b) $$

\n leading to
$$
\nu(r_{_B},\theta,\phi)=1,\quad \nu^{\prime}(r_{_B},\theta,\phi)=0\eqno(50a) $$
$$
Y_{_B}=Y_{_B}(t)=Y(t,r_{_B},\theta,\phi)=\Phi(t,r_{_B}),\quad \left[\frac{ 
Y^{\prime}}{Y}\right]_{_B}=\left[\frac{\Phi^{\prime}}{\Phi}\right]_{_B}  \eqno(50b)
$$
$$
\left[\frac{\p
\nu}{\p\theta}\right]_{r_{_B}}=\left[\frac{\p \nu}{\p\phi}\right]_{r_{_B}}=0,\qquad 
\left[\frac{\p Y}{\p\theta}\right]_{r_{_B}}=\left[\frac{\p Y}{\p\phi}\right]_{r_{_B}}=0
\eqno(50c)
$$

\n where $[]_{r_{_B}}$ means evaluation along $r=r_{_B}$. Therefore, if conditions (49) 
hold, equations (50) guarantee that the hypersurface marked by $(t,r_{_B},\theta,\phi)$ is
a world tube corresponding to the proper time evolution ($t$ is proper time) of a class of
2-spheres labelled by $r=r_{_B}$ , with proper radius $Y_{_B}$. Notice that this
hypersurface has spherical geometry even if the geometry of the regions $0<r<r_{_B}$ and
$r>r_{_B}$ is not spherically symmetric.

Assuming (49) and (50) to hold, we can consider the hypersurface $r=r_{_B}$ as a spherical 
``boundary'' separating two regions: (I) and (II) respectively marked by the ranges:
$0<r<r_{_B}$ and $r>r_{_B}$. A situation of physical interest follows by looking for an
``exterior'' solution for (46), which means considering the system formed by: (I), the
general solution described by (46), the ``interior'', and: for (II), a suitable spherically
symmetric spacetime, the ``exterior''. Since each region (I) and (II) is a chunk of a
different spacetime, we have actualy the problem of a matching between two spacetimes. The
conditions for smoothness of such matchings are the continuity along the matching boundary
of the metric and the extrinsic curvature ($K_{ab}=-n_{a;c}h^c_b$, where $n^a$ is a unit
vector normal to the matching surface). If the same coordinates $ (t,r,\theta,\phi)$ are
used for regions (I) and (II), these matching conditions are $^{[27]}$

$$
\left[g^{^{(I)}}_{ij}-g^{^{(II)}}_{ij}\right]_{_B}=0,\quad \left[K^{^{(I)}}_{ij}-K^{^
{(II)}}_{ij}\right]_{_B}=0,\qquad K_{ij}=\frac{ -g_{ij,r}}{2\sqrt{g_{rr}}} \eqno(51)
$$

\n where $(i,j)=(t,\theta,\phi)$ and $[ ]_{_B}$ means evaluation at $r=r_{_B} $. It  is
straightforward to prove that if the exterior spacetime is a particular case of (46) then
(49)-(50) are equivalent to (51). We shall consider below the cases when region (II) is
either one of: (a) a FLRW subcase, (b) a spherically symmetric but non-FLRW subcase, and
(c) Vaidya spacetime.

\subsection{Matching with a FLRW or a spherical subcase.} 

If region (I) is characterized by a general case of (46) where $A,B_1,B_2,C$ comply with 
boundary conditions (49), then (50) and (51) also hold. If for the whole of region (II) we
have: $n_0=n_0(r_{_B})\equiv\bar n_0$ and $ T_0=T_0(r_{_B})\equiv\bar T_0$, where $\bar
n_0,\bar T_0$ are positive constants, then region (I) is smoothly matched to a FLRW
cosmology. However, the latter is not an arbitrary FRLW cosmology, but the particular case
characterized by (48). It is possible to reverse the roles so that a chunk of a FLRW
spacetime and the general solution (46) respectively occupy regions (I) and (II). Matchings
of this type, especially those in which the FLRW spacetime is the exterior, are
theoretically interesting as models of inhomogeneities within a homogeneous background. The
possibility of performing such matchings has been reported$^{[25]}$ for the perfect fluid
Szekeres-Szafron solution (re-interpreted as a mixture). The result presented here is a
direct generalization to a viscous source. 

It is also possible to match a general solution (46) with spherically symmetric (but 
non-FLRW) subcase, occupying either region (I) or (II). Again, boundary conditions
(49)-(51) must be satisfied and (47) must hold in the region occupied by the spherical
subcase of (46). However, now $ T_0=T_0(r)$ and $n_0=n_0(r)$ in that region. 

\subsection{Matching with a Vaidya exterior.} 

We shall examine the matching, along $r=r_{_{B}}$, between a general solution (the 
interior) and the Vaidya solution (the exterior). The Vaidya spacetime is known$^{[11,12]}$
to be the exterior solution for spherically symmetric imperfect fluid sources and it is
regarded as a radiating generalization of Schwarzschild, since its source is a null dust
(or pure radiation) type of electromagnetic field. It is usually described by the metric
$$
ds^2=-\left[1-\frac{2\tilde m(v)}{Y} \right]c^2dv^2-2c\,dv\,dY+Y^2\left[d \theta^2+
\sin^2\theta d\phi^2 \right]\eqno(52) $$

\n where the null coordinate $v$ is a ``retarded time''. Let the interior be a general 
solution characterized by (46) with $A,B_1,B_2,C$ and $n_0,T_0$ satisfying the boundary
conditions (49)-(50) or (51) at the boundary $ r=r_{_{B}}$, so that this boundary is a
spherical region and the exterior $ r>r_{_{B}}$ is characterized by

$$
ds^2=-c^2dt^2+ \frac{(Y^{\prime})^2}{1-K(r)}dr^2+Y^2\left[d\theta^2+\sin^2 \theta  d\phi^2
\right]\eqno(53)
$$

\n where $Y=Y(t,r)$ (or, equivalently: $Y=\Phi$) and $K$ is given by (47b). If $Y$  in (53)
satisfies (21) with

$$
M=M_{_{B}}\equiv\int_{r=0}^{r_{_{B}}}{n_0Y_0^2Y^{\prime}_0dr},\quad U_0=U_{0_{_{B}}}
\equiv\int_{r=0}^{r_{_{B}}}{\frac{n_0Y_0^2Y^{\prime}_0dr}{ \gamma-1}} ,\quad
Y_0=Y_{0_{_{B}}} \eqno(54a) $$

\n then this exterior is the specific case of Vaidya's spacetime (in the coordinates 
$t,r,\theta,\phi$) with

$$
2\tilde m(v)=\kappa\left[M_{_{B}}+ U_{0_{_{B}}}\left(\frac{Y_{0_{_{B}}}}{ Y_{_{B}}}
\right)^{3(\gamma-1)}\right] \eqno(54b) $$

\n where $Y_{_{B}}=Y(t,r_{_{B}})$, $Y_{0_{_{B}}}=Y_0(r_{_{B}})$, and the relation  between
$v$, $t$ and $Y_{_{B}}$ at the boundary is given by 

$$
\left[\frac{dv}{dt}\right]_{_{B}}=\frac{\pm 1}{\sqrt{1-K_{_{B}}c^2}\pm \dot Y_{_{B}}} 
\eqno(55a)
$$

$$
\left[\frac{dv}{dY}\right]_{_{B}}=\frac{\pm 1}{\dot Y_{_{B}}\left[\sqrt{1-K_{_{B}}c^2}\pm
\dot Y_{_{B}}\right]} \eqno(55b) $$

\n where $\dot Y_{_{B}}$ is (21) with $Y=Y_{_B}$ and $M,U_0,Y_0$ given by (54a), or 
equivalently: $\dot Y_{_{B}}=2\tilde m/Y_{_{B}}-K_{_B}c^2$. The integration of (55b)
provides the trajectory of the boundary in the $(v,Y)$ plane of Vaidya spacetime (52). For
$r>r_{_{B}}$, the transformation relating the coordinates $(t,r)$ in (53) with $(v,Y)$ in
(52) are 

$$
\dot Y dt+ Y^{\prime}dr=dY\eqno(56a)
$$
$$
\frac{cdt}{\sqrt{1-Kc^2}\pm \dot Y}\pm \frac{Y^{\prime}dr}{\sqrt{1-Kc^2}
\left[\sqrt{1-Kc^2}\pm \dot Y\right]}=cdv\eqno(56b) $$

\n where $\dot Y$ is (21) with $M,U_0,Y_0$ given by (54a) but $K=K(r)$ arbitrary.  By
evaluating the corresponding mixed derivatives, it is possible to verify that (56) are well
defined coordinate transformations. 

An interesting result is the fact that the the matching between a general solution and 
Vaidya spacetime is possible, without assuming that the interior region is spherically
symmetric. This occurs also$^{[28]}$ in the matching between Szekeres dust solution and
Schwarzschild (the particular case $U_0=0$ of the matching examined here). Another
interesting point follows from (54b): the time dependent ``Schwarzschild mass'' contains
the contribution from the rest mass ($M$) and internal energy ($U$) of the spherical region
inside the boundary. If this boundary is expanding, so that: $Y_{_{B}}/Y_{0_{_{B}}}\gg 1$,
this mass function decreases tending asymptotically to a constant Schwarzschild mass, and
indicating that the internal energy contribution to the total mass-energy of the spherical
interior is radiated away until only the rest mass contribution remains. 

\section{Concluding remarks.}

We have found a new class of exact solutions of Einstein's field equations, characterized 
by the Szekeres-Szafron metric of class I, whose source is a fluid with shear viscosity.
The solutions admit, in general, no isometries and generalize well known solutions with
dust and perfect fluid sources. However, their main appeal stems from the fact that the
equilibrium state variables (equations (22)-(24)) satisfy a physically meaningful equation
of state: that of a relativistic ideal gas in the following important limits: (1)
ultra-relativistic, (2) non-relativistic and (3) a binary ``matter-radiation'' mixture of
gases of the type (1) and (2). Since the process of integration of the field equations does
not involve the shear viscous stress (equations (11) and (25)), we have provided the
conditions constraining this quantity, and relating it to equilibrium variables and to an
entropy production law, in terms of of the equations of Causal Irreversible Thermodynamics
with phenomenological coefficients related to those obtained from Kinetic Theory for the
Maxwell-Boltzmann gas and the ``radiative gas''. Together with the conditions for
thermodynamic consistency (equations (41)), we have given a regularity condition that
prevents the emergence of unphysical singularities (equation (42)), as well as examined the
fulfilment of the weak, dominant and strong energy conditions (equations (43)-(45)).

The solutions become fully determined once the spatial gradients $ Y^{\prime}/Y$ are 
explicitely computed from the integrals (31)-(33) of the quadrature (30) that follows from
the Friedmann-like evolution equation (21). The temperature and particle number density,
$n_0$ and $T_0$, at an initial hypersurface marked by an arbitrary $t=t_0$, together with a
function related to the extrinsic curvature of this hypersurface ($K$), emerge naturally as
initial value functions. Once a specific solution is determined for any given form of these
functions, it is necessary to test the fulfilment of regularity and energy conditions, as
well as the conditions for thermodynamic consistency. This task is beyond the scope of the
present paper and is already being undertaken in separate papers, dealing with the
classical ideal gas$^{[10]}$ and the matter-radiation mixture $^{[23]}$.

We have also looked at the matching conditions with FLRW and Vaidya spacetimes along a 
spherical boundary. Such a boundary can always be defined, even if the enclosed
``interior'' region is not spherically symmetric. The case of matching to a Vaidya exterior
metric deserves especial attention, since this solution is the usual ``exterior'' metric of
localized (spherical) objects whose sources are imperfect fluids, and there is a large body
of literature concerned with these models. These solutions are, either shear-free fluids
with heat conduction$^{[29]}$, or fluids with anisotropic pressures$^{[30]}$ (but without
associating this anisotropy with viscosity). In many cases, especially those without shear,
these imperfect fluids do not obey physically meaningful equations of state. The imperfect
fluid sources in this paper, having a consistent thermodynamical interpretation, allow one
to examine the geometry and dynamics of a more realistic source expanding or collapsing in
a Vaidya spacetime. The introduction of thermal phenomena and radiative processses into a
collapsing/expanding localized source might lead to interesting generalizations of the so
called ``Swiss Cheese'' models (dust sources matched to vacuoles with Schwarzschild
geometry, see chapter 3 of reference [11]).

Those specific solutions complying with regularity and energy conditions and with 
thermodynamic consistency have an enormous potential as models in applications of
astrophysical and cosmological interest. Consider, for example, the following
possibilities: (1) {\it Cosmological voids}. There is a large body of literature using LTB
(Lem\^aitre-Tolman-Bondi) dust solutions to model the evolution of great voids (see chapter
3 of [11] for a review). Since the solutions in this paper generalize LTB dust solutions,
allowing for a consistent decription of thermal phenomena, it should be possible to
generalize previous work in order to be able to consider those cases in which the presence
of thermal motions (i.e. associated to some kind of dark matter) yields a non-vanishing
pressure. (2) {\it Structure formation in the acoustic phase}. Again, there is a large body
of literature on the study of acoustic perturbations (see for example reference [13]) in
relation to the Jeans mass of surviving cosmological condensations. Equations of state
identical to (5b) are often suggested in this constext$^{[3],[4],[16]}$. Since practically
all work on this topic has been carried on with perturbations on a FLRW background, the
exact solutions derived and presented here, especially the matter-radiation mixture, are
idealy suited as an alternative treatment for this problem. (3) {\it Effect of
inhomogeneities on the microwave cosmic background}. Recent papers$^{[31]}$ have applied
numerical techniques (in the non-linear regime) to explore possible anisotropies in the
microwave background due to photons crossing inhomogeneities, the latter being modelled by
a LTB dust solution. It is worthwhile considering the extension of this work to encompass a
solution whose source can be a thermodynamically consistent photon gas or mixture of a
photon gas and a non-relativistic ideal gas. These and other applications are worth to be
undertaken in future research efforts.

\section{Acknowledgments}

Thus work has been partially supported by the Spanish and Mexican Ministries of Education 
under grants PB94-0718 and CONACYT-3567E.\\ 

\n{\bf{References.}}\\

\n [1] S.R. de Groot, W.A. van Leeuwen and Ch.G. van Weert, {\it  Relativistic Kinetic 
Theory. Principles and Applications}, North Holland Publishing Company, 1980. See pp 46-55.

\n [2] E.W. Kolb and M.S. Turner, {\it The Early Universe}, Addison-Wesley Publishing 
Company, 1990.

\n [3] J. Narlikar, {\it Introduction to Cosmology}, Jones and Bartlett publishers, Inc., 
1983.

\n [4] S. Weinberg {\it Cosmology and Gravitation}, John Wiley and sons, 1971.

\n [5] J. Bardeen, {\it Phys. Rev. D}, {\bf 22}, 1882, (1980); G. F. R. Ellis and M. Bruni, 
{\it Phys. Rev. D}, {\bf 40}, 1804, (1989). 

\n [6] F. Reif, {\it Fundaments of Statistical and Thermal Physics}, McGraw-Hill, 
Kogakusha Ltd. 1965.

\n [7] D. Jou, J. Casas V\'azqez, G. Lebon. {\it {Extended Irreversible Thermodynamics.}}. 
Springer Verlag, Berlin, Heidelberg, New York, 1996. 

\n [8] W. Israel and I. Stewart, {\it {Ann. Phys. (NY).}}, {\bf 118}, 341, (1979).  See
also: W. Israel, in {\it {Relativistic Fluid Dynamics.}} Eds. M. Anile and Y.
Choquet-Bruhat, Spriger-Berlin, 1989. 

\n [9] W. Hiscock and L. Lindblom, {\it {Ann. Phys. (NY).}}, {\bf 151}, 466, (1983).

\n [10] R.A. Sussman, {\it Class. Quantum Grav.}, {\bf 15}, 1759, (1998). 

\n [11] A. Krasi\'nski. {\it {Inhomogeneous cosmological models}}. Cambridge University 
Press, Cambridge, U.K., 1997. 

\n [12] D. Kramer, H. Stephani, M.A.H. MacCallum, E. Herlt. {\it {Exact solutions of 
Einstein's field equations}}. Cambridge University Press, Cambridge, U.K., 1980.

\n [13] R. Maartens and J. Triginer, {\it Pys. Rev. D}, {\bf 56}, 4640, (1997);  ``Acoustic
oscillations and viscosity••, to be published in {\it Phys. Rev. D}, (1998).

\n [14] P. Szekeres, {\it Commun. Math. Phys.}, {\bf 41}, 55, (1975). 

\n [15] D.A. Szafron, {\it J. Math. Phys.}, {\bf 18}, 1673, (1977). 

\n [16] S. Weinberg, {\it Ap. J.}, {\bf 168}, 175, (1971). 

\n [17] P.T. Landsberg and D. Evans, {\it Mathematical Cosmology}, Oxford University 
Press, 1979. (see p. 138).

\n [18] H. Quevedo and R.A. Sussman, {\it Class. Quantum. Grav.}, {\bf 12}, 859, (1995).

\n [19] A. Krasi\'nski, H. Quevedo and R.A. Sussman, {\it J. Math. Phys.}, {\bf 38},  2602,
(1997).

\n [20] P.F. Byrd and M.D. Friedman, {\it Handbook of Elliptic Integrals for Engineers  and
Physicists}, Springer-Verlag, 1954. 

\n [21] W.A. Hiscock and J. Salmonson 1991, {\it {Phys. Rev. D}}, {\bf 43}, 3249.

\n [22] R. Maartens 1995, {\it {Class. Quantum Grav.}}, {\bf 12}, 1455. 

\n [23] D. Pav\'on and R.A. Sussman, ``Adiabatic but irreversible mixtures of matter  and
radiation with thermodynamical consistency,'' (1998). In preparation.

\n [24] S.W. Goode and J. Wainwright, {\it {Phys. Rev. D}}, {\bf 26}, 3315, (1982).

\n [25] R.A. Sussman, {\it {Class. Quantum Grav.}}, {\bf 9}, 1891, (1992). 

\n [26] M.O. Calv\~ao and J.A.S. Lima, {\it {Phys. Lett. A}}, {\bf 141}, 229, (1989).  See
also: J. Garecki and J. Stelmach, {\it {Annals of Physics}} , {\bf 204}, 315, (1990).

\n [27] H. Stephani, {\it General Relativity, An introduction to the theory of the 
gravitational field.}, Cambridge University Press, 1982. 

\n [28] W.B. Bonnor, {\it Commun. Math. Phys.}, {\bf 51}, 191, (1976). 

\n [29] W. B. Bonnor, A.K.G. de Oliveira and N.O. Santos, {\it Phys Rep.}, {\bf 181},  no.
5, 269, (1989).

\n [30] L. Herrera and J. Mart\'{\i}nez, {\it Class. Quantum Grav.}, {\bf 15} , 407, 
(1998).

\n [31] J.V. Arnau, M.J. Fullana, L. Monreal and D. S\'aez, {\it Ap. J.}, {\bf 402},  359,
(1993).

\end{document}